\documentclass{article}

\usepackage[dblblindworkshop,final]{neurips_2025}
\usepackage{graphicx}
\usepackage{subcaption}
\workshoptitle{MLForSys}

\usepackage[utf8]{inputenc} 
\usepackage[T1]{fontenc}    
\usepackage{hyperref}       
\usepackage{url}            
\usepackage{booktabs}       
\usepackage{amsfonts}       
\usepackage{nicefrac}       
\usepackage{microtype}      
\usepackage{xcolor}         
\usepackage{graphicx}
\usepackage{wrapfig}
\usepackage{subcaption} 

\title{When to Reason: Semantic Router for vLLM}

\author{%
  Chen Wang \\
  IBM Research\\
  Yorktown Heights, NY, 10598 \\
  \texttt{Chen.Wang1@ibm.com} \\
  \And
  Xunzhuo Liu \\
  Tencent \\
  \texttt{bitliu@tencent.com} \\
  \And
  Yuhan Liu\\
  University of Chicago \\
  \texttt{yuhanl@uchicago.edu} \\
  \AND
  Yue Zhu \\
  IBM Research \\
  Yorktown Heights, NY, 10598 \\
  \texttt{yue.zhu@ibm.com} \\
  \And
  Xiangxi Mo \\
  UC Berkeley \\
  \texttt{xmo@berkeley.edu}
  \And
  Junchen Jiang\\
  University of Chicago \\
  \texttt{junchenj@uchicago.edu} \\
  \And
  Huamin Chen\\
  Red Hat \\
  Boston, MA, 02210 \\
  \texttt{hchen@redhat.com} \\
}

\begin{document}

\maketitle


\begin{abstract}
Large Language Models (LLMs) demonstrate substantial accuracy gains when augmented with reasoning modes such as chain-of-thought and inference-time scaling. However, reasoning also incurs significant costs in inference latency and token usage, with environmental and financial impacts, which are unnecessary for many simple prompts. We present a semantic router that classifies queries based on their reasoning requirements and selectively applies reasoning only when beneficial. Our approach achieves a $10.2$ percentage point improvement in accuracy on the MMLU-Pro benchmark while reducing response latency by $47.1\%$ and token consumption by $48.5\%$ compared to direct inference with vLLM. These results demonstrate that semantic routing offers an effective mechanism for striking a balance between accuracy and efficiency in open-source LLM serving systems.
\end{abstract}

\section{Introduction}

Large Language Models (LLMs) achieve notable accuracy gains when augmented with advanced inference techniques such as chain-of-thought reasoning or inference-time scaling. Yet, these benefits come at substantial computational and energy costs, particularly when reasoning is applied indiscriminately. Prior studies \cite{wilhelm2025beyond} show that while reasoning improves performance on complex tasks, it is unnecessary for many straightforward queries. This tension makes selective reasoning a central challenge for practical LLM systems.

Recent frameworks such as LangChain/LangGraph \cite{langchain2025routing} and DSPy \cite{khattab2024dspy} enable modular routing policies, but they require manual configuration and are tied to higher-level stacks. In contrast, open-source inference engines like vLLM \cite{kwon2023efficient}—the de facto standard for high-throughput LLM serving—deliver efficient inference but lack native semantic routing. Related systems (e.g., llm-d \cite{llm-d2025}, Production Stack \cite{productionstack2025}) provide lightweight routing but do not support fine-grained control over reasoning. Consequently, developers using vLLM’s APIs avoid vendor lock-in but remain without integrated mechanisms for adaptive reasoning.

To address this gap, we propose a semantic router for open-source inference engines. Our system integrates with vLLM and cloud-native routing frameworks (Envoy, ext\_proc), classifies queries by intent, and selectively applies reasoning only when beneficial. Experiments on the MMLU-Pro benchmark across 14 domains show that our router achieves higher accuracy while reducing latency and token usage by nearly half.

Our contributions are as follows:
\begin{itemize}
    \item We identify the need for semantic routing in open-source inference engines to enable reasoning-aware inference.  
    \item We design, implement, and open-source \cite{vllmsemanticrouter2025} a high-performance and scalable semantic router that integrates with vLLM and Envoy/ext\_proc for fine-grained reasoning control, accelerating Cloud Native ecosystem integration.
    \item We evaluate the semantic router on the MMLU-Pro benchmark and show that it improves accuracy by $10.2$ percentage points while reducing response latency by $47.1\%$ and token consumption by $48.5\%$ compared to direct vLLM inference.
\end{itemize}

\section{Background}
\label{background}

\subsection{Routers in LLM Systems}
Recent work has explored the use of routers to improve the efficiency and accuracy of LLM inference by dynamically deciding how queries should be handled. FrugalGPT \cite{chen2023frugalgpt} achieves up to 98\% cost reduction by learning which combinations of LLMs to invoke for different queries, leveraging prompt adaptation, approximation, and cascaded model selection across commercial APIs. RouteLLM \cite{ongroutellm} similarly trains router models to choose between stronger and weaker LLMs during inference, guided by human preference data and augmentation, yielding substantial cost savings while maintaining accuracy across benchmarks such as MT Bench, MMLU, and GSM8K. These approaches highlight the promise of router-based techniques for improving inference performance, but they remain focused on model-level routing.

\subsection{The Need for Selective Reasoning}
While advanced reasoning strategies such as Chain-of-Thought (CoT) prompting can improve accuracy, recent studies highlight that reasoning is not universally beneficial and often incurs substantial computational overhead. Wilhelm et al. \cite{chen2023frugalgpt} demonstrate that CoT can increase energy costs by up to 150 times while offering little benefit for knowledge-based tasks. Similarly, Aggarwal et al. find that LLMs frequently “overthink” simple queries and “underthink” complex ones \cite{aggarwal2025optimalthinkingbench}, leading to inefficiencies. Meta-analyses by Sprague et al. \cite{sprague2024cot} and the original CoT work by Wei et al. \cite{wei2022chain} further establish that CoT primarily improves performance on math and logic tasks, with limited gains elsewhere and even degraded accuracy in smaller models. To mitigate these inefficiencies, recent frameworks \cite{chen2025aware,zhu2025think,wei2025stop} introduce adaptive reasoning strategies that dynamically regulate reasoning depth, reducing token usage while maintaining accuracy.

\subsection{Semantic Routing}
A semantic router refers to an emerging class of request forwarding systems for LLM inference, in which routing decisions are guided by the semantic meaning of the input rather than by explicit keywords or manually defined rules \cite{manias2024semantic,aurelio2025semanticrouter}. The router operates by encoding both user queries and candidate routing utterances into high-dimensional embeddings \cite{zhang2025query} that capture contextual meaning, and then selecting the target pathway with the highest semantic similarity, typically measured using metrics such as cosine distance. Semantic routing provides a lightweight and efficient mechanism for query-level control, making it a promising foundation for reasoning-aware routing.

\section{System Design}
\label{system-design}
\subsection{System Design}
\label{sys-design-arch}

\begin{figure*}[t]
    \centering
    \begin{subfigure}[t]{0.23\textwidth}
        \centering
        \includegraphics[width=\linewidth]{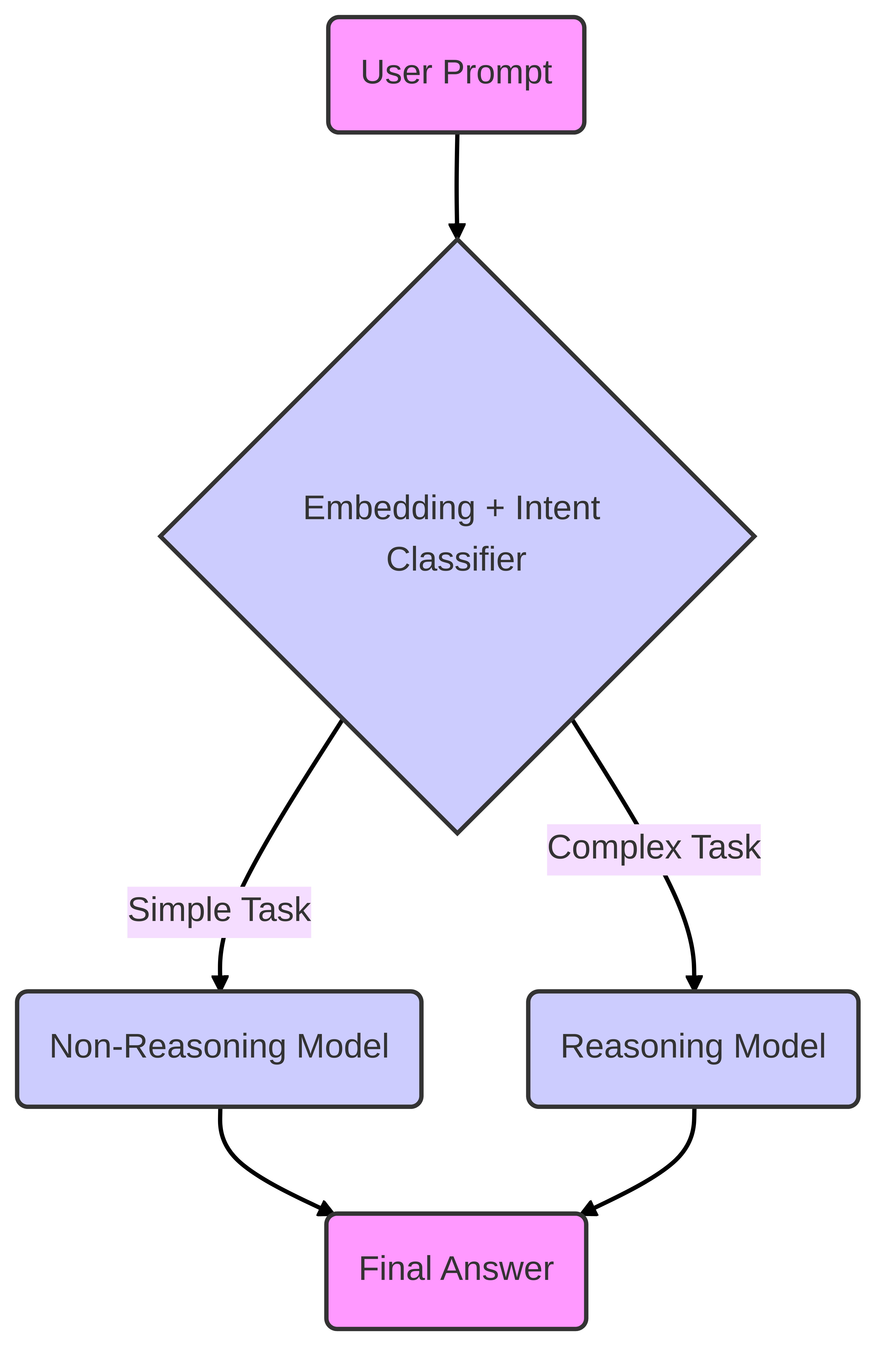}
        \caption{Workflow}
        \label{fig:router-flow}
    \end{subfigure}
    \hfill
    \begin{subfigure}[t]{0.76\textwidth}
        \centering
        \includegraphics[width=\linewidth]{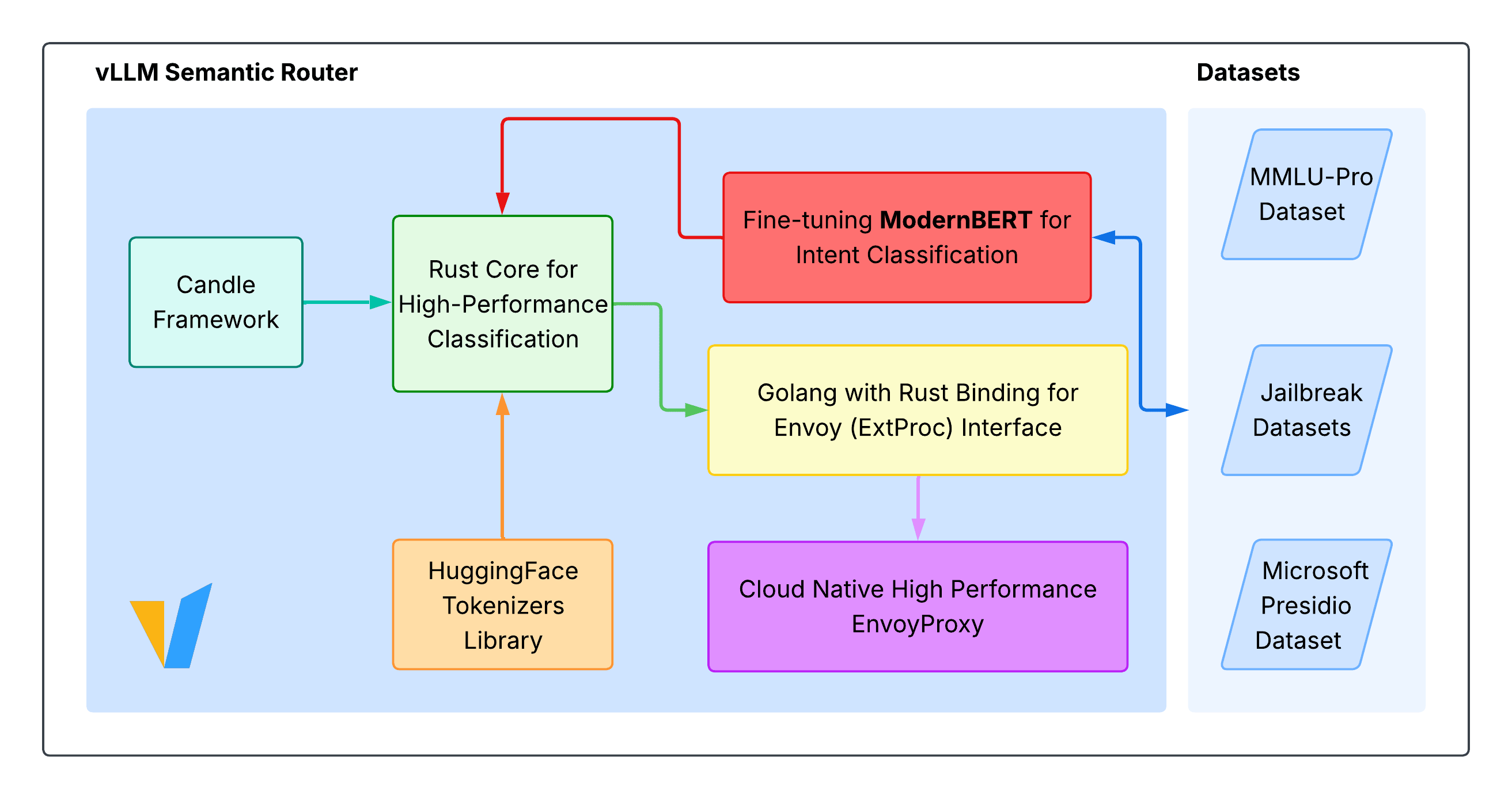}
        \caption{System architecture}
        \label{fig:router-arch}
    \end{subfigure}
    \caption{Overview of the proposed intent-aware semantic router. (a) Workflow of classification and routing; (b) system architecture.}
    \label{fig:router-overview}
\end{figure*}

Our system integrates a semantic router with a reasoning mode selector to dynamically balance efficiency and accuracy in LLM inference. As shown in Figure~\ref{fig:router-flow}, the process begins by encoding the user prompt into high-dimensional semantic embeddings, which capture the contextual meaning of the input. These embeddings are then processed by an intent classifier that determines whether the prompt corresponds to a simple factual query or a reasoning-intensive task. Based on this classification, the router directs the input to the most suitable inference pathway: lightweight inference with a non-reasoning model for simple tasks, or reasoning inference with a chain-of-thought–enabled model for complex queries. Finally, the outputs are unified into a final response. Unlike prior router approaches such as FrugalGPT and RouteLLM, which primarily operate at the model-selection level to trade off accuracy and cost, our design focuses on semantic intent–based routing and selectively invoking reasoning. This enables adaptive reasoning where costly step-by-step inference is applied only when beneficial, while maintaining low latency and efficiency for straightforward queries.

\subsection{Implementation}
The implementation of our intent-aware semantic router integrates three key modules—ModernBERT fine-tuning for intent classification, a Rust-based high-performance classification core, and Golang–Rust bindings for Envoy integration—into a unified architecture, as illustrated in Figure~\ref{fig:router-arch}.
\subsubsection{ModernBERT Fine-tuning for Intent Classification}
We fine‑tune ModernBERT \cite{modernbert-warner2024smarter} —fast,  memory-efficient, supports long contexts, and achieves high accuracy by incorporating modern LLM innovations like RoPE and FlashAttention—for multi‑task intent classification. The training pipeline ingests three datasets: MMLU‑Pro \cite{wang2024mmlu} (\textasciitilde 12K academic samples across \textasciitilde 14 domains), Microsoft Presidio \cite{microsoftpresidioresearch} (\textasciitilde 50K token‑level PII examples), and jailbreak security datasets \cite{chao2024jailbreakbench}. The classification pipeline can use either CPU or GPU for real-time inline inference and simplifies the runtime environment resource requirements.

\subsubsection{Rust Core for High-Performance Classification}
The classification engine is implemented in Rust using Hugging Face’s Candle framework \cite{huggingfacecandle}, which enables efficient, zero-copy tensor workflows, SIMD acceleration, and optimized memory usage. It runs multi‑stage parallel inference—category classification, PII detection, and jailbreak detection—leveraging Rust’s ownership model for thread safety. The pipeline batches requests and utilizes Hugging Face Tokenizers for fast tokenization, supports large context window, and chains multiple classification tasks, sustaining highly concurrent requests on commodity hardware without using expensive GPUs.

\subsubsection{Golang + Rust (via CGO) for Cloud-Native Envoy Integration}
We wrap the Rust-based classification core in a Golang layer using CGO bindings to support Envoy’s External Processing (ext\_proc) filter interface \cite{envoyextproc}. Envoy intercepts HTTP requests and forwards them via gRPC to the external processor, which applies real‑time classification and routing decisions before responses reach backend services. The CGO layer is statically linked, minimizing runtime overhead while enabling seamless integration with Kubernetes, service meshes, and API gateway patterns. Such design pattern facilitates Cloud Native ecosystem adoption.

\section{Evaluation}
\label{sec:results}

We evaluate our semantic router on an NVIDIA L4 GPU using the Qwen/Qwen3-30B-A3B model served by vLLM v0.10.1 with tensor parallelism degree 4. The evaluation is conducted on the MMLU-Pro benchmark across 14 domains, measuring accuracy, token usage, and latency. For direct vLLM comparison, we run the same model under six execution modes—neutral reasoning (NR) and explicit chain-of-thought (XC), each with reasoning enabled or disabled configurations.

\begin{figure}[htbp]
  \centering
  \includegraphics[width=0.9\linewidth]{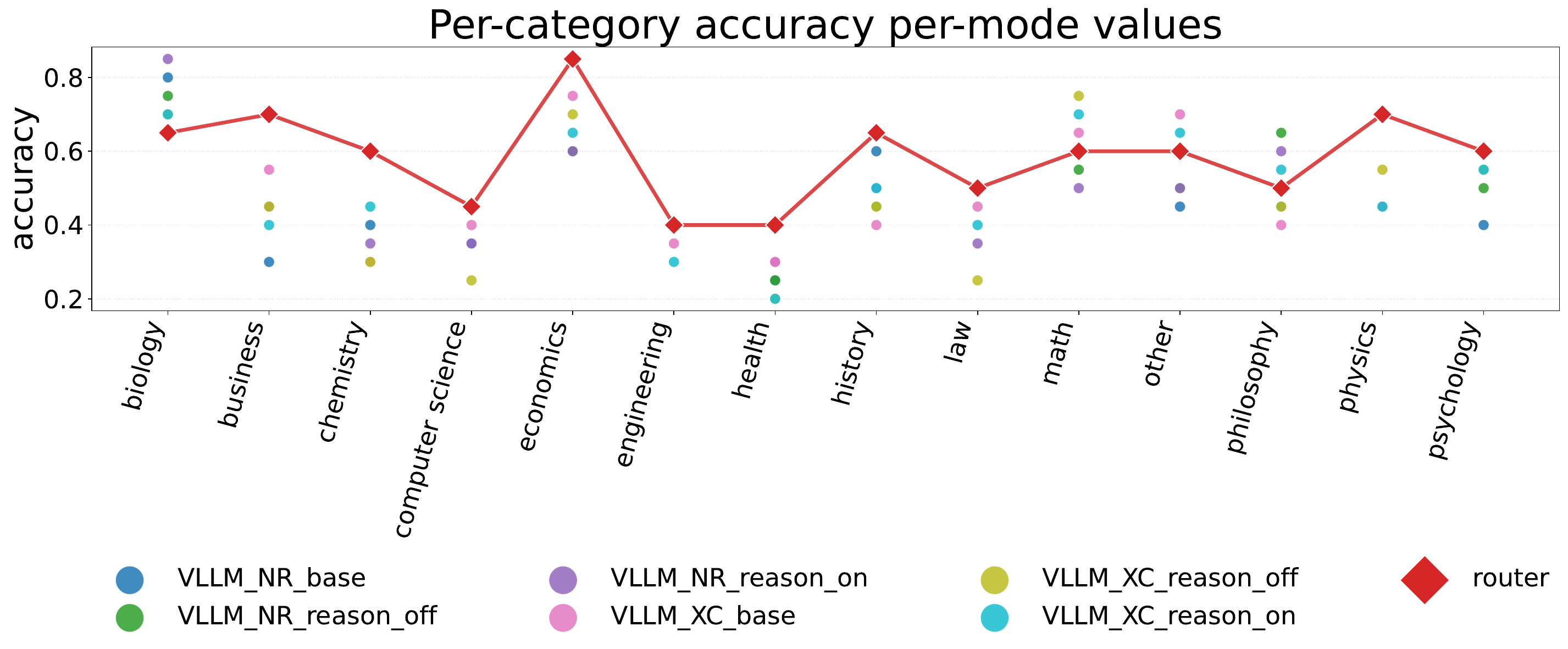}
  \caption{Per-category accuracy across 14 MMLU-Pro domains for direct vLLM modes and our semantic router.}
  \label{fig:per_category_accuracy}
\end{figure}

Figure~\ref{fig:per_category_accuracy} breaks down accuracy by the 14 MMLU-Pro domains for all execution modes (NR/XC with \texttt{reason\_on}, \texttt{reason\_off}, and \texttt{base}), along with our semantic router. Across the majority of categories, the router leads in reasoning-heavy domains and remains competitive in knowledge-centric areas, indicating that selective reasoning does not sacrifice accuracy on fact-focused tasks while delivering benefits where structured reasoning is essential.

Table~\ref{tab:overall_performance} summarizes the aggregate performance metrics comparing our semantic router against direct vLLM inference. Overall, the semantic router improves accuracy by 10.24 points while cutting latency by 47.1\% and token usage by 48.5\%.

\begin{table}[htbp]
\centering
\caption{Overall performance of semantic router versus direct vLLM inference on MMLU-Pro.}
\label{tab:overall_performance}
\begin{tabular}{lccc}
\toprule
\textbf{Method} & \textbf{Avg. Accuracy} & \textbf{Avg. Latency (s)} & \textbf{Avg. Tokens} \\
\midrule
Semantic Router & \textbf{58.57\%} & \textbf{13.09} & \textbf{887.5} \\
Direct vLLM & 48.33\% & 24.76 & 1,722.1 \\
\midrule
\textbf{Improvement} & \textbf{+10.24pp} & \textbf{-47.1\%} & \textbf{-48.5\%} \\
\bottomrule
\end{tabular}
\end{table}

Our evaluation shows that the semantic router delivers substantial efficiency gains while improving overall accuracy, achieving a statistically significant 10.24 percentage point increase (p < 0.01) with 48.5\% fewer tokens and 47.1\% lower latency. The router is particularly effective in knowledge-intensive domains such as business and economics, where accuracy improvements exceed 20 percentage points, while performance in technical areas like engineering and computer science remains more challenging. Mixed results in reasoning-heavy domains (e.g., mathematics and biology) highlight opportunities for refining routing strategies. Overall, the router demonstrates robust improvements across 11 of 14 domains, underscoring its ability to match queries to appropriate reasoning strategies. These results suggest that semantic routing offers a practical path toward more accurate and cost-efficient LLM inference in production settings.

\section{Conclusion}
This paper presented a semantic router that dynamically selects between reasoning and non-reasoning strategies to optimize large language model inference. Evaluation on MMLU-Pro shows that the router improves accuracy by more than 10 percentage points while reducing token usage and latency by nearly 50\%. The approach is particularly effective in knowledge-intensive domains such as business, economics, and physics, though challenges remain in technical and reasoning-heavy areas. Integrated with vLLM, the router demonstrates that semantic routing is a practical and efficient solution for real-world inference serving.

\medskip
{
\small
\bibliographystyle{plain}  
\bibliography{references}  
}


\appendix

\section*{Appendix A. Additional Per-Category Results}

\begin{figure}[htbp]
    \centering
    \includegraphics[width=\linewidth]{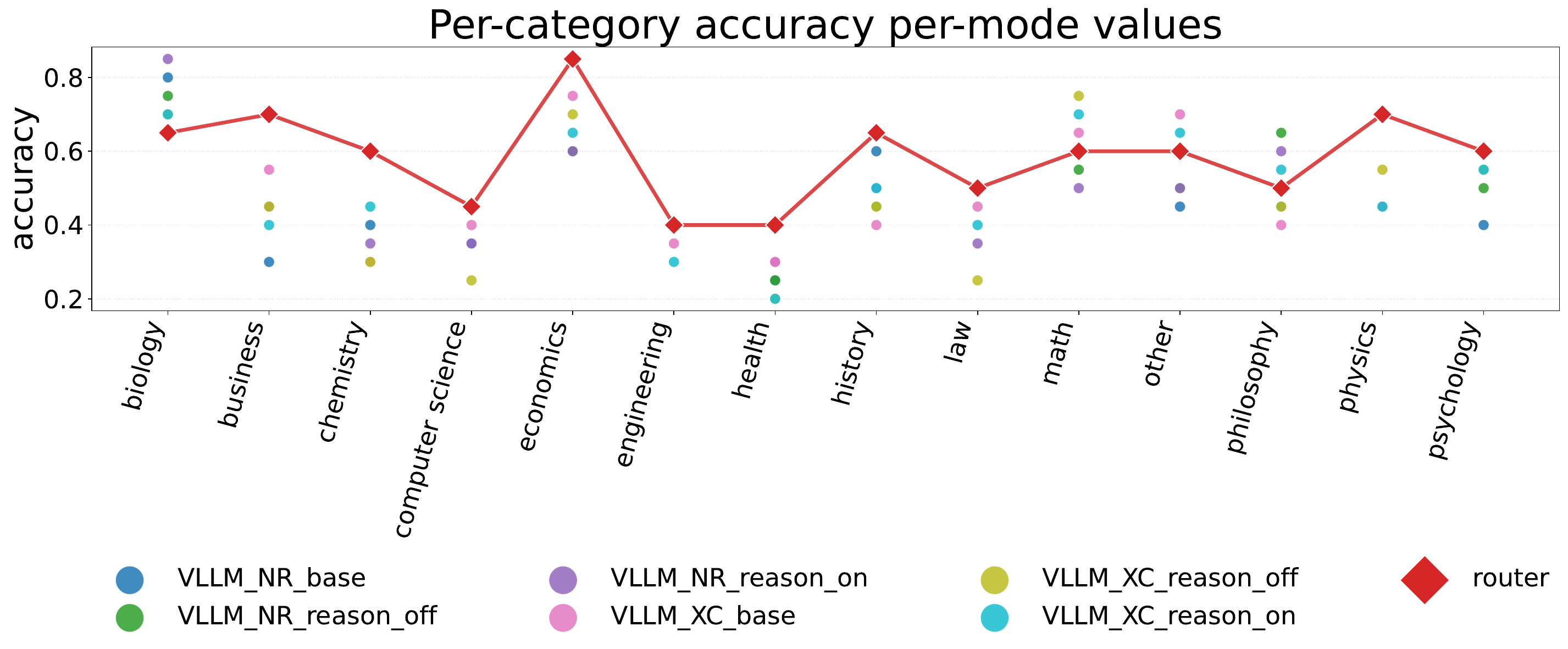}
    \caption{Per-category accuracy across all inference modes on MMLU-Pro.}
    \label{fig:accuracy}
\end{figure}

In addition to the per-category accuracy results reported in Figure~\ref{fig:accuracy}, we include
two supplementary breakdowns that highlight the efficiency benefits of semantic routing.

\begin{figure}[htbp]
    \centering
    \includegraphics[width=\linewidth]{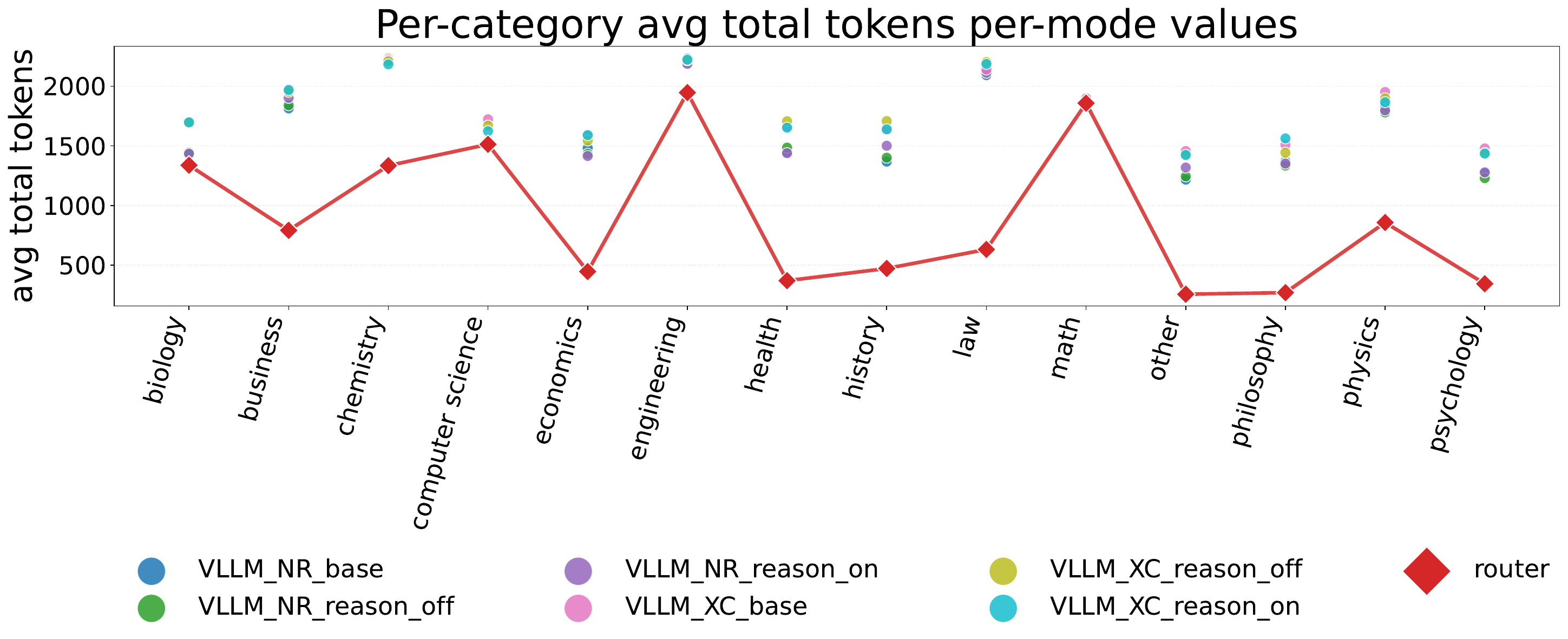}
    \caption{Per-category average total tokens across all inference modes on MMLU-Pro. 
    The semantic router consistently achieves the lowest token usage, reducing overhead 
    in knowledge-centric domains (e.g., history, law, health) while remaining competitive 
    in reasoning-heavy areas such as math and physics.}
    \label{fig:tokens}
\end{figure}

\begin{figure}[htbp]
    \centering
    \includegraphics[width=\linewidth]{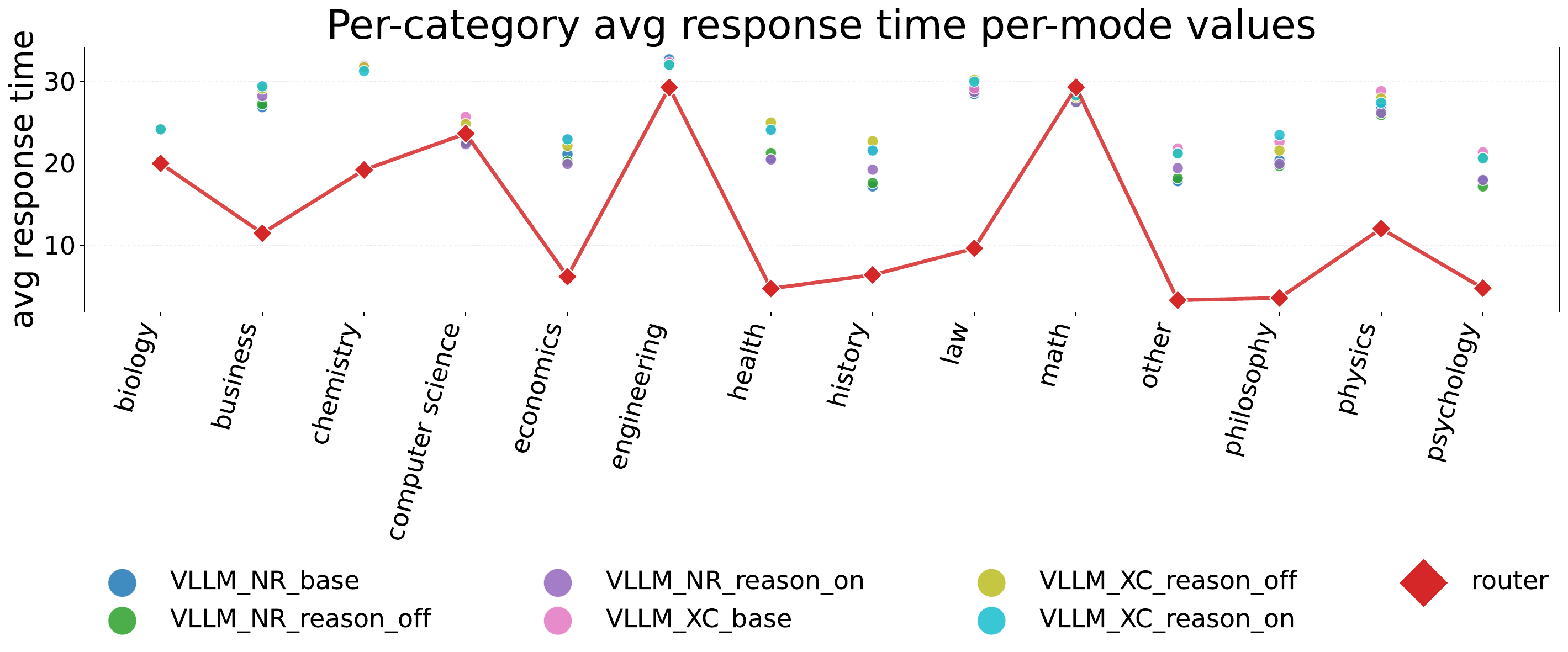}
    \caption{Per-category average response latency across all inference modes on MMLU-Pro. 
    The semantic router reduces latency substantially compared to direct vLLM modes, 
    particularly in domains with shorter factual queries (e.g., history, philosophy). 
    Even in complex reasoning categories, the router sustains lower response times by 
    avoiding unnecessary reasoning overhead.}
    \label{fig:latency}
\end{figure}

The per-category breakdowns in Figures~\ref{fig:tokens} and~\ref{fig:latency} confirm that 
the semantic router consistently improves efficiency across domains. In terms of token usage, 
the router reduces average consumption by nearly half relative to direct vLLM execution modes, 
with especially pronounced savings in knowledge-intensive subjects such as history, law, and health, 
where reasoning is rarely required. Similarly, the latency results show that the router sustains 
substantially faster response times across most categories, cutting delays by over 40\% even in 
reasoning-sensitive areas like mathematics and physics. These results demonstrate that semantic 
routing not only improves aggregate efficiency but also achieves robust per-domain benefits, 
delivering faster and cheaper inference without sacrificing accuracy.
\end{document}